\documentclass[pdflatex,sn-nature]{sn-jnl}

\usepackage{graphicx}%
\usepackage{multirow}%
\usepackage{amsmath,amssymb,amsfonts}%
\usepackage{amsthm}%
\usepackage[title]{appendix}%
\usepackage{xcolor}%
\usepackage{textcomp}%
\usepackage{manyfoot}%
\usepackage{booktabs}%
\usepackage{algorithm}%
\usepackage{algorithmicx}%
\usepackage{algpseudocode}%
\usepackage{listings}%
\usepackage{gensymb}%
\usepackage{siunitx}

\usepackage{tabularx}
\usepackage{bm} 
\usepackage{comment}
\usepackage{dcolumn}
\usepackage{makecell, tabularx} 
\setcellgapes{3pt} 

\theoremstyle{thmstyleone}%
\theoremstyle{thmstyletwo}%

\theoremstyle{thmstylethree}%

\begin{document}

\title[Cluster-configurational study of G-center in Silicon]{Cluster-configurational study of G-center in Silicon}

\author[1]{\fnm{Narayan} \sur{Pokhrel}}
\author[1,2]{\fnm{Kyungwha} \sur{Park}}
\author*[1,2,3]{\fnm{Vsevolod} \sur{Ivanov}}\email{vivanov@vt.edu}

\affil[1]{\orgdiv{Department of Physics}, \orgname{Virginia Tech}, \orgaddress{\city{Blacksburg}, \postcode{24061}, \state{Virginia}, \country{USA}}}
\affil[2]{\orgname{Virginia Tech Center for Quantum Information Science and Engineering}, \orgaddress{\city{Blacksburg}, \postcode{24061}, \state{Virginia}, \country{USA}}}
\affil[3]{\orgname{Virginia Tech National Security Institute}, \orgaddress{\city{Blacksburg}, \postcode{24060}, \state{Virginia}, \country{USA}}}

\abstract{Understanding the properties of defects is imperative for proper use for variety of applications including quantum computing. In this paper, we use the multiconfigurational self consistent field (MCSCF) combined with DFT optimized geometry in order to investigate the spin and optical properties of G centers in Silicon. By utilizing quantum chemistry based methods, we show excellent agreement with the Zero Phonon Line and Zero Field Splitting Tensor components of the G center. We also calculate the theoretical spin decoherence time of the G centers using Cluster Correlation Expansion (CCE) methods.}


\maketitle

\section{Introduction}
Optically active point defects in semiconductors are promising platforms for quantum technologies due to their flexibility and ease of integration with existing wafer-scale fabrication pipelines for on-chip optical and electronic components \cite{Quard2025_integration}. As a result, they are actively being developed for a number of applications, including efficient single photon emitters \cite{Aharonovich2016_spe_review}, quantum sensors \cite{Degen2017_Quantum_Sensing_RMP}, precision magnetometers \cite{Aslam2023_biomedical, Manas-Valero2025_NVmagnetometry}, quantum communication \cite{Atature2018_quantum_app_review, Hensen2015_NV_Bell}, as well as local strain sensors for host material damage by high energy particles \cite{ang2026multiscalereconstructionsingleiondamage, araujo2025nuclearrecoildetectioncolor} and radiation \cite{PhysRevApplied.20.014058}. Furthermore, color center defects can be engineered to have application-specific properties \cite{defects_by_design}, which has driven the creation of large-scale defect databases \cite{Xiong2023_siliconDB, DAVIDSSON2021108091, ivanov2023databasesemiconductorpointdefectproperties}.

Of the host materials being explored, silicon is particularly promising due to its widespread use in microelectronics \cite{Xiong2024, Day2024, Filippatos_TMIGW}. Several defect centers in silicon have emerged as promising quantum platforms, including the T-center, a spin $S=1/2$ defect with nuclear spin memory \cite{Johnston2024_Tcenter, Song2026_Tcenter, PhysRevMaterials.6.L053201}, C-center, which has L-band emission \cite{Udvarhelyi2022_Ccenter}, and C$_i$ center which can be programmably erased \cite{Jhuria2024_programmable}. Out of all the known color center defects in silicon, the G-center has received the most attention \cite{Day2024, Redjem2020, Zhiyenbayev2023_scalable, Redjem2023_allsilicon, Redjem2023_defect_engineering} due to its ease of synthesis and emission in the telecom band, as well as its optically addressable metastable electron-spin triplet state \cite{dreau2026_new2026ODMR, ODONNELL1983258}.

\begin{figure}[htbp]
    \centering
    \includegraphics[width=\textwidth]{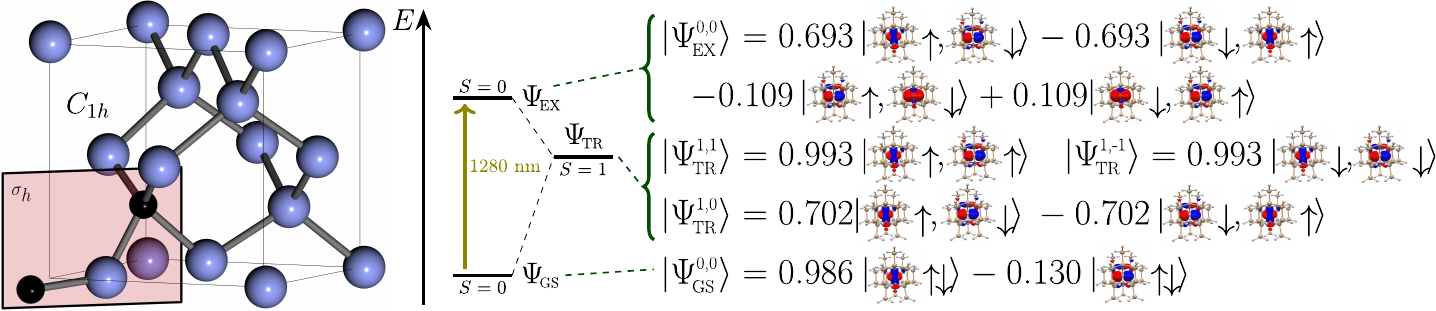}
    \caption{Overview of the G-center structure and electronic states. The G-center structure has $(C_{1h})$ symmetry consisting of two substitutinal carbon atoms (black spheres) in silicon (blue spheres). The electronic structure consists of a singlet ($S=0$) ground and excited states and a metastable triplet ($S=1$) state. Primary multiconfigurational components of each state are shown in terms of the localized orbital basis. } 
    \label{fig:overview}
\end{figure}

The G-center defect structure consists of two carbon atoms, and can take on two possible configurations: Type A (GCA) involving on substitutional C$_s$ and one interstitial C$_i$ carbon, and Type B (GCB), in which the C$_i$ displaces an adjacent silicon to form a C$_s$-Si$_i$-C$_s$ chain \cite{Ivanov2022}. 
The defect has monoclinic $(C_{1h})$ symmetry \cite{foy1981uniaxial}, with a $\sigma_h$ mirror plane coincident with the C$_s$-Si$_i$-C$_s$ chain (Figure \ref{fig:overview}). While the GCA configuration is optically dark, the GCB configuration exhibits an optically bright emission to its singlet ground state with an experimentally measured zero phonon line (ZPL) energy of 0.969 eV \cite{bean1970electron,thonke1981new}, which is well within the optic-fiber telecomunications band. Previous density functional theory (DFT) calculations have found a metastable triplet that lies below the excited singlet state \cite{Ivanov2022,udvarhelyi2021identification}, which has been experimentally probed through optically detected magnetic resonance (ODMR) \cite{lee1982optical, ODONNELL1983258, dreau2026_new2026ODMR}. Additionally, GCB displays a dynamical hopping of its center of mass \cite{durand2024hopping}, which leads to a further splitting of its photoluminescence spectrum. 

There are two major motivations to perform multiconfigurational study on G center defects. First, we derive similarity from NV centers where the single configurational wavefunction fails to replicate the zero phonon line and even the ordering of states \cite{bhandari2021multiconfigurational,benedek2025accurate}. 
While DFT values have been able to reproduce the ZPL energy of the G-center with some accuracy, the single configurational treatment may not capture the full extent of localization and many-electron physics. Namely, the nature of the G-center optical excitation remains unclear; localized excitation, bound-exciton, and exchange-driven continuum mechanisms have been proposed \cite{komza2024indistinguishable, Hong_2026_Gcenter_bound}.  The other motivation is to build a clear picture of the singlet-triplet manifold and higher lying states in the G-center. Recent ODMR measurements of the metastable triplet state have revealed that the triplet sublevels undergo different relaxation pathways \cite{dreau2026_new2026ODMR}, motivating the need for theoretical understanding of spin non-conserving processes. In this work, we present the first comprehensive multiconfigurational study of the G-center optical, spin, and electronic properties. 


\section{Results}
\subsection{Choice of Active Space and Roots}

\begin{figure}[htbp]
    \centering
    \includegraphics[width=0.5\textwidth]{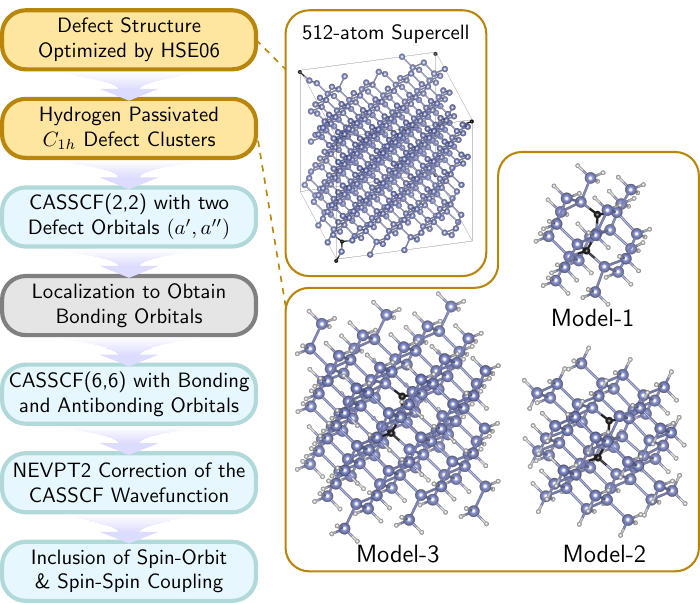}
    \caption{Overview of the computational workflow. The G-center structure is relaxed within a 512-atom supercell using a hybrid functional approach. From this structure, hydrogen-passivated structures are obtained: Model 1 (69 atoms), Model 2 (141 atoms), and Model 3 (265 atoms). Electronic structures for these models are obtained using a 2-orbital + 2-electron CASSCF (2,2) calculation from which the six-orbital basis is obtained through a Pipek-Mezey localization procedure. These orbitals are then used for the full-scale 6-electron + 6-orbital  CASSCF (6,6) calculation, followed by NEVPT2 corrections and postprocessing to obtain spin properties.} 
    \label{fig:flowchart}
\end{figure}

In order to accurately determine the ZPL energy, multireference ab initio method are applied to DFT relaxed geometries \cite{janicka2022computational} . However, such energy falls short of the exact ZPL if the geometry is not accurate. While geometry relaxation for NV centers using CASSCF have been applied previously \cite{benedek2025accurate}, these steps are computationally challenging due to the need of performing CASSCF calculation for each geometric step, which itself is computationally expensive for larger clusters with larger active space. 

\begin{figure}[htbp]
    \centering
    \includegraphics[width=0.6\textwidth]{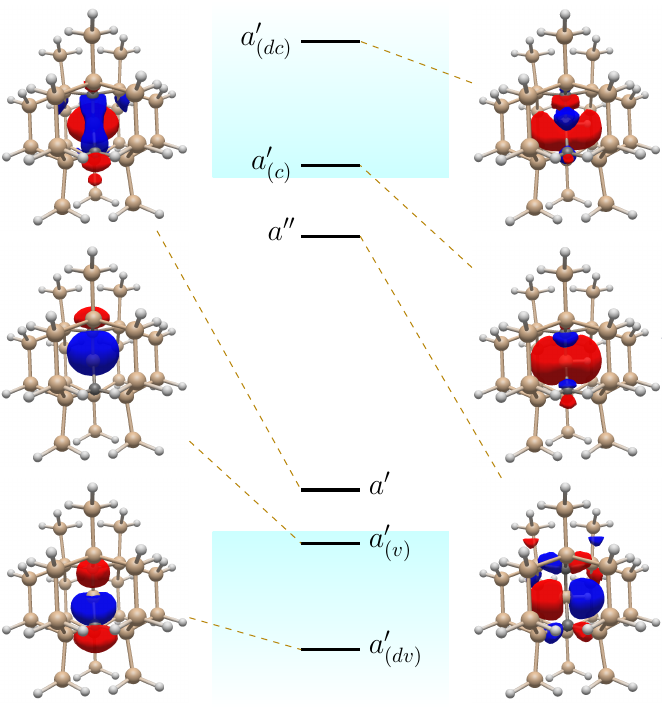}
	\caption{Top view of the 6 defect orbitals in the SA-CASSCF calculation of 69 atom cluster ordered according to energy. Similar 6 active orbitals are present for 141 and 265 atom structure. The AVOGADRO \cite{hanwell2012avogadro} program is used for visualization. 
    }
	\label{fig:orbitals}
\end{figure}

In such scenario, we can calculate the vertical excitation energy in the same ground state geometry using State Average CASSCF (SA-CASSCF). The vertical excitation energy is the energy difference between the excited state and the ground state on same geometry. There are two advantages to performing state average calculation: one, it avoids having to optimize energy and wavefunctions for different states separately, second, the resulting CASSCF calculation has much improved convergence. 

The choice of active space is the crux of quantum chemistry calculations. Here, we present a systematic way to increase the active space for G center configuration. The computational procedure is provided in flowchart Fig \ref{fig:flowchart} . First, we start with two electrons in two orbitals (CAS(2,2)) with the two orbitals being Highest Occupied Molecular Orbital (HOMO) and Lowest Unoccupied Molecular Orbital (LUMO) identified from Hartree-Fock calculations. Both of the orbitals have $sp^3$ character with the HOMO transforming as the function of $ A^{'}$ Irreducible Represenation and LUMO orbital transforming as the function of $ A^{"}$ Irreducible Representation of the $C_{1h}$ point group. This is our minimal active space.

In quantum chemistry method, inclusion of additional orbitals beyond the minimal active space is often required to improve the CASSCF calculations \cite{Roos2016}. Such expansion to the active space is done by identifying defect-localized orbitals in the valence band or orbitals close in energy \cite{bhandari2021multiconfigurational}. However, no such orbitals exist in our virtual space. In order to identify additional orbitals, we use Pipek-Mezey (PM) localization scheme \cite{pipek1989fast}. The PM localization yields two bonding orbitals in the inactive space. With the inclusion of these two bonding orbitals and their antibonding counterparts from virtual space, we perform a second set of CASSCF calculation with 6 electrons and 6 orbitals (CAS(6,6)). 
Both the bonding and antibonding orbitals transform under the A' irreducible representation. So we have 5 A' and 1 A" orbitals.  

\begin{table}[htbp]
\centering
\label{tab:Example}
\caption{Many body states of the G center with 6 electrons and 6 orbitals in terms of Slater Determinant for 265 atom cluster. The states are ordered according to energy with the first state being the lowest in energy. S refers to Singlet and T refers to Triplet. 2,u,d,0 refer to the orbital being  occupied doubly, occupied only by up spin,occupied only by down spin and empty respectively . Only leading states are shown $(>0.1)$}
\begin{tabular}{ c | c | >{\centering\arraybackslash}m{1.0cm} | >{\centering\arraybackslash}p{8.6cm} }
\hline\hline
 & E (eV) & $S, m_S$ &  Configuration  \\
\hline
$\Psi_1^S$ & 0 & 0, 0 & 
$
 0.986 |a'_{v}\bar{a}'_{v}a'\bar{a}'\rangle
-0.130 |a'_{v}\bar{a}'_{v}a''\bar{a}''\rangle
$ \\\hline
$\Psi_2^S$ & 1.040 & 0, 0 & 
$
\begin{aligned}
 0.693 |a'_{v}\bar{a}'_{v}a'\bar{a}''\rangle
&-0.693 |a'_{v}\bar{a}'_{v}\bar{a}'a''\rangle 
-0.109 |a'_{v}\bar{a}'_{v}a''\bar{a}'_{c}\rangle \\[-0.3em]
&+0.109 |a'_{v}\bar{a}'_{v}\bar{a}''a'_{c}\rangle
\end{aligned}
$ \\\hline
$\Psi_3^S$ & 1.898 & 0, 0 & 
$
\begin{aligned}
 0.066 |a'_{v}\bar{a}'_{v}a'\bar{a}'\rangle
&-0.618 |a'_{v}\bar{a}'_{v}a'\bar{a}'_{c}\rangle 
+0.618 |a'_{v}\bar{a}'_{v}\bar{a}'a'_{c}\rangle \\[-0.3em]
&+0.461 |a'_{v}\bar{a}'_{v}a''\bar{a}''\rangle
+0.056 |a'_{v}\bar{a}'_{v}a'_{c}\bar{a}'_{c}\rangle
\end{aligned}
$ \\\hline
$\Psi_4^S$ & 1.763 & 0, 0 & 
$
\begin{aligned}
 0.694 |a'_{v}a'\bar{a}'\bar{a}''\rangle
&-0.694 |\bar{a}'_{v}a'\bar{a}'a''\rangle
-0.071 |\bar{a}'_{v}\bar{a}'a''a'_{c}\rangle \\[-0.3em]
&-0.071 |a'_{v}a'\bar{a}''\bar{a}'_{c}\rangle
\end{aligned}
$ \\\hline
$\Psi_5^S$ & 2.067 & 0, 0 & 
$
\begin{aligned}
-0.106&|a'_{v}\bar{a}'_{v}a'\bar{a}'\rangle
-0.318 |a'_{v}\bar{a}'_{v}a'\bar{a}'_{c}\rangle 
+0.318 |a'_{v}\bar{a}'_{v}\bar{a}'a'_{c}\rangle \\[-0.3em]
&\quad-0.863 |a'_{v}\bar{a}'_{v}a''\bar{a}''\rangle
+0.145 |a'_{v}\bar{a}'_{v}a'_{c}\bar{a}'_{c}\rangle
\end{aligned}
$ \\\hline
$\Psi_6^S$ & 2.748 & 0, 0 & 
$
\begin{aligned}
&\phantom{{}+{}}0.488 |a'_{v}a'\bar{a}'\bar{a}'_{c}\rangle
-0.488 |\bar{a}'_{v}a'\bar{a}'a'_{c}\rangle
+0.483 |a'_{v}\bar{a}'a''\bar{a}''\rangle \\[-0.3em]
&-0.483 |\bar{a}'_{v}a'a''\bar{a}''\rangle
-0.083 |a'_{v}\bar{a}'a'_{c}\bar{a}'_{c}\rangle
+0.083 |\bar{a}'_{v}a'a'_{c}\bar{a}'_{c}\rangle \\[-0.3em]
&-0.085 |a'_{v}a''\bar{a}''\bar{a}'_{c}\rangle
+0.085 |\bar{a}'_{v}a''\bar{a}''a'_{c}\rangle
\end{aligned}
$ \\\hline
 & & 1, 1 & 
$
 0.993 |a'_{v}\bar{a}'_{v}a'a''\rangle
$ \\
$\Psi_1^T$ & 0.680 & 1, 0 & 
$
 0.702 |a'_{v}\bar{a}'_{v}a'\bar{a}''\rangle
+0.702 |a'_{v}\bar{a}'_{v}\bar{a}'a''\rangle
$ \\
 & & 1, -1 & 
$
 0.993 |a'_{v}\bar{a}'_{v}\bar{a}'\bar{a}''\rangle
$ \\\hline
 & & 1, 1 & 
$
 0.993 |a'_{v}\bar{a}'_{v}a'a'_{c}\rangle
$ \\
$\Psi_2^T$ & 1.548 & 1, 0 & 
$
 0.702 |a'_{v}\bar{a}'_{v}a'\bar{a}'_{c}\rangle
+0.702 |a'_{v}\bar{a}'_{v}\bar{a}'a'_{c}\rangle
$ \\
 & & 1, -1 & 
$
 0.993 |a'_{v}\bar{a}'_{v}\bar{a}'\bar{a}'_{c}\rangle
$ \\\hline
 & & 1, 1 & 
$
-0.979 |a'_{v}a'\bar{a}'a''\rangle
-0.127 |a'_{v}\bar{a}'a''a'_{c}\rangle
+0.076 |a'_{v}a'\bar{a}''a'_{c}\rangle
$ \\
$\Psi_3^T$ & 1.731 & 1, 0 & 
$
\begin{aligned}
&-0.692 |a'_{v}a'\bar{a}'\bar{a}'_{c}\rangle
-0.692 |\bar{a}'_{v}a'\bar{a}'a'_{c}\rangle
-0.076 |a'_{v}\bar{a}'a''\bar{a}'_{c}\rangle \\[-0.3em]
&+0.076 |\bar{a}'_{v}a'\bar{a}''a'_{c}\rangle
+0.068 |a'_{v}a'\bar{a}''\bar{a}'_{c}\rangle
-0.068 |\bar{a}'_{v}\bar{a}'a''a'_{c}\rangle
\end{aligned}
$ \\
 & & 1, -1 & 
$
-0.979 |\bar{a}'_{v}\bar{a}'a'\bar{a}''\rangle
+0.127 |\bar{a}'_{v}a'\bar{a}''\bar{a}'_{c}\rangle
-0.076 |\bar{a}'_{v}\bar{a}'a''\bar{a}'_{c}\rangle
$ \\\hline
 & & 1, 1 & 
$
\begin{aligned}
 -0.789 |a'_{v}a'a''\bar{a}''\rangle
&-0.589 |a'_{v}a'\bar{a}'a'_{c}\rangle
 +0.094 |a'_{v}a'a'_{c}\bar{a}'_{c}\rangle \\[-0.3em]
&+0.070 |a'_{v}a''\bar{a}''a'_{c}\rangle
\end{aligned}
$ \\
$\Psi_4^T$ & 2.766 & 1, 0 & 
$
\begin{aligned}
&-0.558 |a'_{v}\bar{a}'a''\bar{a}''\rangle
-0.558 |\bar{a}'_{v}a'a''\bar{a}''\rangle
-0.416 |a'_{v}a'\bar{a}'\bar{a}'_{c}\rangle \\[-0.3em]
&-0.416 |\bar{a}'_{v}a'\bar{a}'a'_{c}\rangle
+0.066 |a'_{v}\bar{a}'a'_{c}\bar{a}'_{c}\rangle
+0.066 |\bar{a}'_{v}a'a'_{c}\bar{a}'_{c}\rangle \\[-0.3em]
&\quad+0.050 |a'_{v}a''\bar{a}''\bar{a}'_{c}\rangle
+0.050 |\bar{a}'_{v}a''\bar{a}''a'_{c}\rangle
\end{aligned}
$ \\
 & & 1, -1 & 
$
\begin{aligned}
-0.789 |\bar{a}'_{v}\bar{a}'\bar{a}''a''\rangle
&-0.589 |\bar{a}'_{v}\bar{a}'a'\bar{a}'_{c}\rangle
+0.094 |\bar{a}'_{v}\bar{a}'\bar{a}'_{c}a'_{c}\rangle \\[-0.3em]
&+0.070 |\bar{a}'_{v}\bar{a}''a''\bar{a}'_{c}\rangle
\end{aligned}
$ \\\hline
 & & 1, 1 & 
$
 0.986 |a'_{v}\bar{a}'_{v}a''a'_{c}\rangle
+0.058 |\bar{a}'_{v}a'a''a'_{c}\rangle
$ \\
$\Psi_5^T$ & 2.710 & 1, 0 & 
$
 0.697 |a'_{v}\bar{a}'_{v}a''\bar{a}'_{c}\rangle
+0.697 |a'_{v}\bar{a}'_{v}\bar{a}''a'_{c}\rangle
$ \\
 & & 1, -1 & 
$
 0.986 |a'_{v}\bar{a}'_{v}\bar{a}''\bar{a}'_{c}\rangle
-0.058 |a'_{v}\bar{a}'\bar{a}''\bar{a}'_{c}\rangle
$ \\\hline
 & & 1, 1 & 
$
 0.783 |a'_{v}a'\bar{a}'a'_{c}\rangle
-0.593 |a'_{v}a'a''\bar{a}''\rangle
-0.126 |a'_{v}a''\bar{a}''a'_{c}\rangle
$ \\
$\Psi_6^T$ & 2.566 & 1, 0 & 
$
\begin{aligned}
&\phantom{{}+{}}0.783 |a'_{v}a'\bar{a}'\bar{a}'_{c}\rangle
+0.783 |\bar{a}'_{v}a'\bar{a}'a'_{c}\rangle
-0.419 |a'_{v}\bar{a}'a''\bar{a}''\rangle \\[-0.3em]
&-0.419 |\bar{a}'_{v}a'a''\bar{a}''\rangle
-0.089 |a'_{v}a''\bar{a}''\bar{a}'_{c}\rangle
-0.089 |\bar{a}'_{v}a''\bar{a}''a'_{c}\rangle
\end{aligned}
$ \\
 & & 1, -1 & 
$
 0.783 |\bar{a}'_{v}a'\bar{a}'\bar{a}'_{c}\rangle
-0.593 |\bar{a}'_{v}\bar{a}'a''\bar{a}''\rangle
-0.126 |\bar{a}'_{v}a''\bar{a}''\bar{a}'_{c}\rangle
$ \\\hline
\hline
\end{tabular}
\end{table}

It is necessary to tune the number of roots to be selected for state averaging in CASSCF + NEVPT2 calculation. For the roots, we find that inclusion of larger number of roots decreases the NEVPT2 energy gap between the states, however, \emph{ad hoc} addition of larger number of roots for state average leads to divergence of CASSCF wavefunction and calculated properties. Conversely, very small number of roots can result in poor set of active orbitals since the orbitals are optimized for an incomplete set of electronic states. In order to find the optimal number of roots to include for state average, we start with equal number of roots for singlet and triplet (2 singlet and 2 triplet) and continue adding equal number of roots to both the singlet and triplets on state average until the ZPL energies and ZFS components are converged. We also monitor the active orbitals to see if any orbitals change their characteristics upon addition of higher excited roots. We find that CASSCF calculation with 10 roots (5 singlets+ 5 triplets) leads to a converged estimation of the wavefunction and the Zero Field Splitting parameters for CAS(6,6). 

\begin{table}[htbp]
    \makegapedcells 
    \caption{\label{tab:table1}SA-CASSCF + NEVPT2 Vertical Excitation energy $\Delta E_\text{VE}$ between Ground Spin-Singlet state and the Excited Spin-singlet state. The CASSCF wavefunction is averaged over 10 roots (5 singlets + 5 triplets)}
    \vspace{15pt}
    \begin{tabular}{|c|c|c|c|}
        \hline
        \textbf{basis set} & \textbf{Model 1} & \textbf{Model 2} & \textbf{Model 3} \\
        \hline
         cc-pVDZ-DK & 1.612 &  1.229 & 1.036\\
        cc-pVTZ-DK & 1.568 & 1.174 & \\
        \hline
    \end{tabular}
\end{table}
\subsection{Electronic Structure}
In order to understand the role of considered geometry, we studied the excitation energy levels for both the first excited singlet and triplet.  The transition energies are provided in table \ref{tab:table1}. We observe that when increasing the size from Model-1 to Model-2 the change in energy is considerable. However, on further increasing the model size to Model-3, we find that the energy is convergent within 0.1 eV for both the singlet and triplet. It is worthwhile to note that increasing the basis set produces a very little change in energy ($0<0.05 eV$) so we believe that the 265 atom cluster is a convergent geometry size in order to do fully understand the physics of the G center. 

Vertical excitation energy for the G center with 265 atom cluster is 1.07 eV. This value is close to the experiment Zero Phonon Line  considering the error of 0.1 eV for CASSCF + NEVPT2 calculations \cite{sarkar2022assessing}. Some deviations from experimental ZPL is expected because we are performing the calculation on same ground state singlet geometry. A full treatment of ZPL requires geometry optimization for each cluster, subsequent CASSCF calculations and inclusion of electron-phonon effect, which is beyond the scope of this work.

\begin{table*}[htbp]
\centering
\begin{tabular}{|c|c|c|c|c|c|c|}
\hline
\textbf{ZFS (MHz)} & \multicolumn{3}{c|}{\textbf{CAS(2,2)}}
& \multicolumn{3}{c|}{\textbf{CAS(6,6)}} \\
\textbf{components}
& Model 1 & Model 2 & Model 3
& Model 1 & Model 2 & Model 3\\
\hline
$|D_{xx}|$ & 0.0030 & 0.5996& 0.5796
 & 0.2039  & 0.0909 &  0.0240 \\
$|D_{yy}|$ & 1.0373  & 0.7495& 0.7295
  & 0.8364 & 1.2841 & 1.0613\\
$|D_{zz}|$& 1.0403  & 1.3491 & 1.3091
  &  1.0403  & 1.3750  & 1.0852 \\
\hline
\end{tabular}
\caption{Zero field splitting components of the SOC + SSC hamiltonian of the first excited triplet with CASSCF + NEVPT2. CAS(2,2) is carried out by averaging over all possible configurations (1 singlets + 3 triplets), CAS(6,6) calculation is carried out on 12 states (6 singlets + 6 triplets)}
\label{tab:zfs}

\end{table*}
\subsection{Zero Field Splitting (ZFS)}
All the electronic spin-triplets are split by SOC and dipolar electron spin-spin coupling (SSC). We focus on the level splitting of the ground spin-triplet state which can be described by the following ZFS Hamiltonian.
\begin{equation*}
    \hat H = D\left(S_z^2 - \frac{1}{3} S(S+1)\right) + E(S_x^2-S_y^2)
\end{equation*}
where $D$ and $E$ are uniaxial and transverse ZFS parameters respectively, and $S$ is the pseudospin operator with S=1. For S=1, the eigenvalues of the ZFS Hamiltonian are D + E, D - E and 0. 

Diagonalization of the SOC+SSC Hamiltonian yields a set of spin-orbit-split sublevels. The D tensor of the triplet state is obtained through  the energy difference between the QDPT eigenstates of the SOC+SSC hamiltonian. The traceless tensor components are listed in table \ref{tab:zfs}. Increasing the cluster size leads to significant changes in the predicted ZFS. This sensitivity may reflect the multiconfigurational nature of the defect electronic structure, where the inclusion of additional orbitals can alter the balance of electronic configurations contributing to the wavefunction.




\begin{figure}[htbp]
	\includegraphics[width=1\columnwidth]{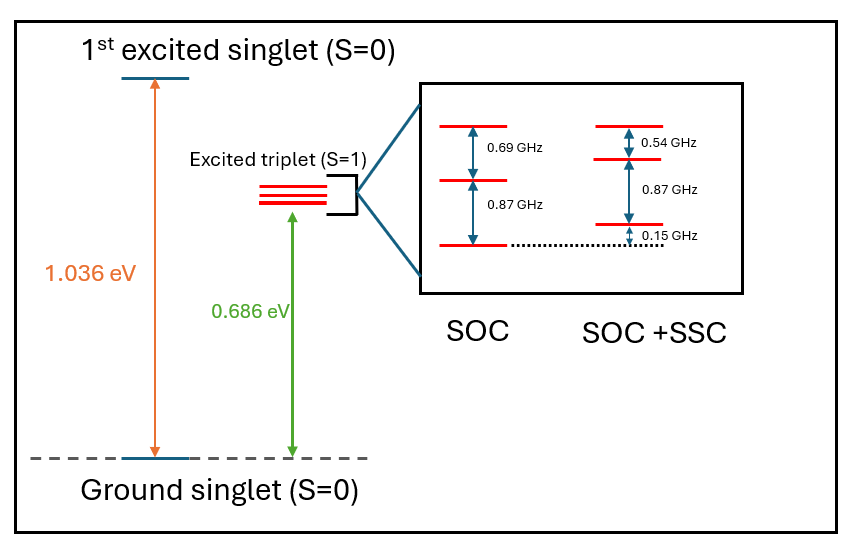}
	\caption{Energy diagram showing the relative position of Singlets and Triplets in SA-CASSCF calculations. The inset show the level splitting of the triplet level by SOC and SOC+SSC}
	\label{qdpt splitting}
\end{figure}

\subsubsection{Intersytem crossings}
Intersystem crossings (ISC) provide a nonradiative pathway for population transfer between electronic states of different spin multiplicity, which would otherwise be forbidden, through the spin orbit coupling. 
The ISC rate between an initial state $|i\rangle$ and a final state $|f\rangle$ can be calculated using Fermi's golden rule:
\begin{equation*}
    k_{i\rightarrow f} = \frac{2\pi}{\hbar} |\langle\Psi_f|\hat H_{SO}|\Psi_i\rangle|^2 \rho(E_f-E_i)
\end{equation*}

where $\rho(E)$ is the vibrational density of states accounting for Franck-Condon overlap and $|\langle\Psi_f|\hat H_{SO}|\Psi_i\rangle|$ is the electronic coupling term evaluated between spin free states. 

The complete calculation of ISC rates requires consideration of direct SOC term and two complex terms arising from spin-vibrational couplings. While such work is outside the scope of the current work, dominant ISC channels can be obtained from calculation of SOC matrix elements. The SOC elements matrix between the triplet excited state and singlet ground state for the CAS(10,8) wavefunction is presented in table \ref{tab:soc}

\begin{table}[ht]
\centering
\caption{\label{tab:socme}Spin-Orbit Coupling (SOC) Matrix Elements}
\begin{tabular}{ccc} 
\toprule
\multicolumn{2}{c}{\textbf{Transition}} & \textbf{SOC Matrix Element} \\ 
\cmidrule(lr){1-2} 
Initial State & Final State & $\langle \Psi_i | \hat{H}_{SO} | \Psi_f \rangle$ (cm$^{-1}$) \\ 
\midrule
$T_0$ & $S_0$ & 62.5 \\
$T_0$ & $S_2$ & 65.4 \\
$T_4$ & $S_4$ & 53.7 \\
$T_3$ & $S_0$ & 49.4 \\
$T_1$ & $S_4$ & 29.8 \\
$T_4$ & $S_2$ & 29.5 \\
$T_0$ & $S_4$ & 27.1 \\
\bottomrule
\end{tabular}
\end{table}

Intersystem crossing was analyzed through the calculated spin–orbit coupling matrix elements between the low-lying singlet and triplet states. Larger coupling strengths and smaller energy gaps were taken as indicators of enhanced nonradiative spin-forbidden transitions.

The triplet state has three sublevels that differ by the projection $m_s$ of the total electron spin. Two of the sublevels of the lowest triplet state are used as qubit states. The intersystem crossing rate is spin dependent, and by optically exciting the G center, one can effectively reset the qubit into $|m_s\rangle = 0$ state within microseconds (needs to be looked at, colloquium paper)
We have a well-separated electronic ground state and a quasi-degenerate excited manifold, where SOC enables selective intersystem crossing, but does not fully mix spin states.Our system is a optically driven defect with a single dominant ISC pathway inside the excited manifold.
Intersystem crossing is mediated by a dominant spin-orbit coupling channel ($\sim$75 cm$^-1$) within the first excited manifold, which lies $\sim$0.95 eV above the ground state. The resulting spin mixing is perturbative but sufficient to enable efficient nonradiative decay pathways. 
The triplet manifold is already strongly split and spin-mixed by SOC. Eigenstates are spin-orbit entangled.

\subsection{Excited State Lifetime}
Recent Experiments have shown the excited lifetime for the G centers to be around $\tau = 4.5- 6$  $ns$ \cite{durand2024genuine, baron2022single, beaufils2018optical, kim2025bright}.  The radiative lifetime in the case of G center can be estimated with the magnitude transition dipole moment (TDM) $(|\vec\mu|)$ for the excitation from ground singlet to excited singlet and the energy difference between the states ($v$) using the Wigner-Weisskopf theory.\cite{weisskopf1930berechnung}

\begin{equation*}
    \frac{1}{\tau} = \frac{n_r (2\pi)^3v^3|\vec{\mu}|^2}{3\epsilon_0 \hbar c^3}
\end{equation*}

The TDM for the G-center with the CASSCF calculations is 3.31 Debye which is in line with the DFT calculated value of ~$2.8$ Debye. \cite{komza2024indistinguishable} The Transition Dipole moment is aligned along the $1\bar10$ axis of the defect.
With the calculated value of the vertical excitation energy (1.07 eV) and TDM, we estimate the radiative lifetime of the excited singlet to be around $ 0.26 \mu s$. This is the theoretical lifetime. 
However, the decay of the excited state may also occur from non-radiative channels affecting the overall radiative lifetime through $\tau_{exp} = 1/(k_r + k_{nr})$, where $k_r$ and $k_{nr}$ are the radiative and non-radiative rates respectively. Quantum efficiency can be estimated as the radiative fraction of the total decay rate, $\Phi = \tau_{exp}/\tau_r$, where $\tau_r$ is the radiative lifetime \cite{savarese2012fluorescence}. Since, G-centers show very little quantum efficiency $<1\%$ on the higher end \cite{durand2024genuine}, this rescaling with respect to the quantum efficiency of the defect gives the expected radiative lifetime $\tau_{exp}$ in the range of $2.6 ns$. 
This is in same order of magnitudes as the experimental lifetime but off by a factor of 2.

\subsection{Coherence Time Measurement}
\begin{figure}[htbp]
    \centering
	\includegraphics[width=0.7\columnwidth]{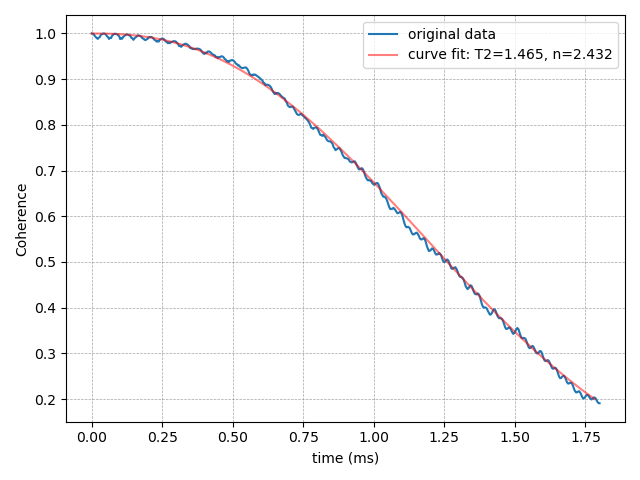}
	\caption{Coherence plot along with the fitted curve}
	\label{fig:coherence}
\end{figure}

Using PYCCE with the converged parameter of $r_{bath} = 26 $nm $r_{dipole} = 6$nm and CCE-order of 2, we obtain the decoherence function of the central spin of G center, as listed in Fig \ref{fig:coherence}. We sample up to 10 different bath states and isotopic configurations in order to obtain an accurate curve for the Coherence function. Convergence of the PYCCE parameters are listed in supplementary materials. 

The decoherence function is fitted with the compressed exponent exp[${-(t/{T_2})^n}]$ and the resulting coherence is obtained to be 1.465 ms. This sugggests that the decoherence mediated through the nuclear spin bath interactions allows for a substantial coherence time, however, experimental lifetime has been subpar, limiting factors include the electronic interactions and the excited lifetime of the metastable triplet sublevels. \cite{cache2026optical}

\section{Discussion}

\begin{table}[htbp]
\makegapedcells
\centering
\label{tab:comparison}
\caption{Comparison of computed \cite{Ivanov2022, komza2024indistinguishable, udvarhelyi2021identification} and experimental \cite{ODONNELL1983258} characteristics of the silicon G-center, including optical properties: zero-phonon line (ZPL) and transition dipole moment (TDM), and spin properties: energy of the lowest energy triplet relative to the ground state and zero-field splitting (ZFS) components. The asterisk* indicates that an open-shell correction was used for the singlet excited state.}
\begin{tabular}{ >{\raggedright\arraybackslash}p{1.9cm} | >{\centering\arraybackslash}p{1.7cm} | >{\centering\arraybackslash}p{1.7cm} | >{\centering\arraybackslash}p{1.7cm} | >{\centering\arraybackslash}p{1.7cm} | >{\centering\arraybackslash}p{1.3cm}  }
\hline\hline
 Property & Ivanov et al. \cite{Ivanov2022} & Komza et al. \cite{komza2024indistinguishable} & Udvarhelyi et al. \cite{udvarhelyi2021identification} & This work& Expt. \cite{ODONNELL1983258}  \\
\hline
Method & HSE 2x2x2 & HSE & HSE+GW & CAS(6,6)& - \\
\# atoms & 216 & 512 & 512 & 265 & - \\
\hline
ZPL (meV)           & 987  & 1000 & 985* & 1040 & 969\\
$E(\Psi_1^T)$ (meV) & 591  &   -  & 678  & 680  &  - \\
TDM$^2$ (D$^2$)     & 4.3  & 7.84 &   -  & 10.95&  - \\
$|D_{xx}|$ (MHz)    & 152  &   -  & 307  &  24  & 142\\
$|D_{yy}|$ (MHz)    & 964  &   -  & 911  & 1061 & 800\\
$|D_{zz}|$ (MHz)    & 1116 &   -  & 1218 & 1085 & 941\\
\hline\hline
\end{tabular}
\end{table}

Inspection of the wavefunction of the states in terms of slater determinant reveals that the while the ground singlet and first excited triplet are dominated by a single configuration, the first excited singlet has a sizable mixture of other configurations. This suggests that while single-electron descriptions of the lowest triplet and singlet states may be qualitatively correct, they may fail to correctly describe the excited singlet due to its multiconfigurational nature. In fact, while previous DFT calculation suggest that the ZPL primarily arises because of the bound exciton transfer of the electron from the valence band to the in gap-state \cite{Ivanov2022}. However, our multiconfigurational calculations support a the picture that the ZPL is corresponds to a transition between two localized orbitals \cite{komza2024indistinguishable}. 

A numerical comparison of our results against previous studies and experiment is shown in Table \ref{tab:comparison}. While the ZPL value of 1040 meV in our work compares well with those obtained using the $\Delta$-SCF method with hybrid functionals, our approach has the advantage of not needing any corrections or tuning parameters. Specifically, HSE06+U calculations were able to find a ZPL of 985 meV \cite{udvarhelyi2021identification}, which is remarkably close to the experimental value, but required expensive GW calculations to obtain the Hubbard-$U$ parameter and an open-shell correction of the excited state energy. A similar value of 987 meV was obtained using a $k$-point resolved $\Delta$-SCF method on a $2\times2\times2$ momentum-space grid \cite{Ivanov2022}, however this approach requires knowledge of the correct occupations at each $k$-point. 

Beyond the ZPL, our work also finds an energy for the spin-triplet state that is in good agreement with the prior HSE06+U study \cite{udvarhelyi2021identification}. The improved localization of the defect states results in a significant enhancement of the TDM, bringing it closer to the expected value from experimental lifetime measurements. The multiconfigurational approach also produces a $D_{zz}$ component that is in better agreement with the experiment, although predictions for the other ZFS components are worse, which may be a limitation of the cluster size.




\section{Methods}

\subsection{Cluster Preparation}

The geometry of the 217-atom G Center cluster optimized with HSE06 hybrid DFT functional as obtained from \cite{Ivanov2022} is used for all the calculations.

First, using the HSE06 optimized geometry, we eliminate atoms starting from outer layer in different steps. With the obtained cluster, we replace the outer silicon with hydrogen with bond length of 1.46 pm and same direction of bond as the original $Si-Si$ bond. No subsequent relaxation step is carried out.

The above process applied on the 217 atom G center file gives three different clusters: first is a larger cluster with 265 atoms  (C$_2$Si$_{125}$H$_{138}$) and second is a medium sized cluster with 141 atoms (C$_2$Si$_{61}$H$_{78}$)  and an even smaller cluster with 69 atoms (C$_2$Si$_{25}$H$_{42}$). The 69, 141, and 256 atoms clusters are referred to as Model-1, Model-2 and Model-3 respectively in the remainder of this paper. All the three clusters retain the $C_s$ symmetry of the defect.

\subsection{Quantum Chemistry}
The quantum chemistry method is carried out in three steps: (i) Complete Active Space Self Consistent Field (CASSCF) calculation with state averaging (SA). (ii) N Electron Valence Perturbation Theory (NEVTP2) correction to the SA-CASSCF energies. (iii) Inclusion of Spin Orbit Coupling (SOC) and Spin Spin Coupling (SSC). 

The multireference ab initio calculations are performed without enforcing any symmetry (C1) using the ORCA\cite{ORCA}\cite{ORCA5} code (ORCA 6.0.1). Scalar relativistic effects are included based on the second-order Douglas-Kroll-Hess Hamiltonian and relativistically contracted all-electron correlation-consistent (cc) basis sets.\cite{de2001parallel}\cite{dunning1989gaussian} 
RIJCOSX approximation \cite{neese2009efficient} is used to speed up the calculation with def2/J \cite{weigend2006accurate} auxiliary basis set for coulomb  integrals and pVDZ/C \cite{weigend2002efficient} auxiliary basis set for correlation.
For the 69 atoms and 141 atoms, we use both the polarized double $\zeta$ (cc-pVDZ-DK) and polarized triple $\zeta$ (cc-pVTZ-DK) contraction for all Carbon, Silicon and Hydrogen atoms, while for the 256 atom cluster we use only the cc-pVDZ-DK basis set because of the computational limit. The effects of different basis sets on different cluster structures are present in the Results section.

With the aim of reducing the computational demand for NEVPT2 calculations, we use domain based local pair of natural orbitals (DLPNO) NEVPT2. \cite{guo2016sparsemaps}. Applying DLPNO-NEVPT2 on the small cluster showed similar results for transition energies. Spin-Orbit Coupling (SOC) and Spin-Spin Coupling (SSC) are introduced through Quasi Degenerate Perturbation Theory (QDPT) as implemented in ORCA. \cite{ganyushin2006first} 

\subsection{Decoherence Dynamics}

Spin-bath induced decoherence of the central spin is measured with the coherence function. Using Cluster Correlation Expansion(CCE) method, the coherence function can be reduced into a factor of irreducible contributions from the spin-bath clusters. \cite{onizhuk2021pycce}

\begin{equation*}
    \mathcal{L}(t) = \prod_{c=1}^n \tilde L_c
\end{equation*}
where $\tilde L_c$ is the contribution from bath cluster upto $c$ spins. The maximum cluster included $n$ determines the order of the CCE expansion.

With the aim of understanding the spin coherence in G centers interacting with the nuclear spin bath, we use first-principles generalized cluster correlation expansion (gCCE) calculations with Monte Carlo bath state sampling using the PYCCE package \cite{onizhuk2021pycce}. Using the 512 atom DFT optimized cluster, we calculate the electron-nuclear hyperfine coupling of the cluster containing the G center with the HSE06 functional in VASP. Using DFT-computed hyperfine parameters of the 512 atom and point-dipole approximation for the outer bath spins, we then calculate the interactions between the Silicon interstitial and nuclear spins in the host matrix.

\bibliography{references}

\section{Acknowledgements}

VI acknowledges support from Virginia Tech startup funds. The authors acknowledge Advanced Research Computing at Virginia Tech (arc.vt.edu) for providing computational resources and technical support. VI acknowledges support from the National Science Foundation Growing Convergence Research Award 2428507.

\section*{Author Contributions}
Narayan Pokhrel performed all theoretical calculations with input from Vsevolod Ivanov and Kyungwha Park. Vsevolod Ivanov and Kyungwha Park provided overall supervision of the project. Narayan Pokhrel and Vsevolod Ivanov drafted and edited all parts of the manuscript. The work was conceived with contributions from all authors.
\section*{Competing Interests}
The authors declare no competing interests.
\section*{Data Availability}
The data that support the findings of this study are available from the corresponding author upon reasonable request.
\section*{Code Availability}
The code used to produce the results is available from the corresponding author upon reasonable request.

\end{document}